\documentclass[preprint]{aastex}
\usepackage{times}
\newcommand{\hide}[1]{} 

\usepackage{amsmath}

\usepackage[draft,textsize=scriptsize,textwidth=1.1in,linecolor=black,backgroundcolor=yellow]{todonotes}
\usepackage{soul,color}
\newcommand{\sje}[2][]{\textcolor{red}{\st{#1}}\textcolor{blue}{#2}}
\newcommand{\sjc}[2][]{\hl{#1}\todo[inline]{#2}\noindent}
\newcommand{\sjd}[3][]{\textcolor{red}{\st{#1}}\textcolor{blue}{\hl{#2}}\todo[inline]{#3}\noindent}
\soulregister{\cite}{7}
\soulregister{\citep}{7}
\soulregister{\citet}{7}
\soulregister{\ref}{7}
\soulregister{\sje}{7}
\soulregister{\sjc}{7}
\soulregister{\sjd}{7}
 
\begin{document}

\title{\textbf{\LARGE Formation and Evolution of Binary Asteroids}}
\author {\textbf{\large Kevin J. Walsh}}
\affil{\textbf{\emph{\small\em Southwest Research Institute}}}
\author {\textbf{\large Seth A. Jacobson}}
\affil{\textbf{\emph{\small\em  Observatoire de la Cote d'Azur and University of Bayreuth}}}
\begin{abstract}
\begin{list}{ } {\rightmargin 1in}
\baselineskip = 11pt
\parindent=1pc {\small Satellites of asteroids have been discovered in
  nearly every known small body population, and a remarkable aspect of
  the known satellites is the diversity of their properties.  They
  tell a story of vast differences in formation and evolution
  mechanisms that act as a function of size, distance from the Sun,
  and the properties of their nebular environment at the beginning of
  Solar System history and their dynamical environment over the next
  4.5 Gyr. The mere existence of these systems provides a laboratory
  to study numerous types of physical processes acting on asteroids
  and their dynamics provide a valuable probe of their physical
  properties otherwise possible only with spacecraft.

Advances in understanding the formation and evolution of binary
systems have been assisted by: 1) the growing catalog of known
systems, increasing from 33 to $\sim$250 between the Merline et al.
(2002) {\it Asteroids III} chapter and now, 2) the detailed study and
long-term monitoring of individual systems such as 1999 KW$_4$ and
1996 FG$_3$, 3) the discovery of new binary system morphologies and
triple systems, 4) and the discovery of unbound systems that appear to
be end-states of binary dynamical evolutionary paths.

Specifically for small bodies (diameter smaller than $10$~km), these
observations and discoveries have motivated theoretical work finding
that thermal forces can efficiently drive the rotational disruption of
small asteroids.  Long-term monitoring has allowed studies to
constrain the system's dynamical evolution by the combination of
tides, thermal forces and rigid body physics. The outliers and split
pairs have pushed the theoretical work to explore a wide range of
evolutionary end-states.

~\\~\\~\\~}

\end{list}
\end{abstract}  

\section{\textbf{INTRODUCTION}}
There have been considerable advances in the understanding of the
formation and evolution of binary systems since the {\it Asteroids
  III} review by~\citet{Merline:2002vn} and another comprehensive
review by~\cite{Richardson:2006fo}.  The current properties of this
population are detailed in the chapter by Margot et al., and this
review will rely on their analysis in many places as we review work on
the formation and dynamics of these systems.  While the inventory of
known binary systems in all populations has increased, for some
populations the understanding of dynamics and formation have advanced
only minimally while in others places research has moved rapidly.
Therefore this chapter will not be evenly weighted between different
populations, rather there will be substantial discussion of small
asteroids and the YORP effect.

The scope of this chapter will be different than the {\it Asteroids
  III} chapter by~\citet{Merline:2002vn}.  Thanks to an excellent
review of binary systems in the Kuiper Belt in the {\it Solar System
  Beyond Neptune} book by~\citet{Noll:2008ku}, and the apparent
physical, dynamical and evolutionary differences between binary minor
planets in the outer and inner regions of the Solar System, we will
exclude the Kuiper Belt population from our discussion here.

\bigskip
\noindent
\textbf{ 1.1 The Known Population of Binary Minor Planets}
\bigskip

The known population of asteroid satellites has increased from 33 to
$\sim$244 between the Merline et al.  2002 {\it Asteroids III} chapter
and now: 49 near-Earth, 19 Mars-crossing, and 93 Main Belt asteroids
~\citep{Pravec:2007fw,Johnston:2014wm}.  As noted in 2002, the binary
systems found among near-Earth asteroids (NEAs) have only a subset of
the properties of those found among the Main Belt asteroids
(MBAs). While the orbital and collisional dynamics differ
substantially in these two populations, further study has found that
the variation and similarities between binary properties is most
strongly dependent on size.

\begin{figure*}
 \plotone{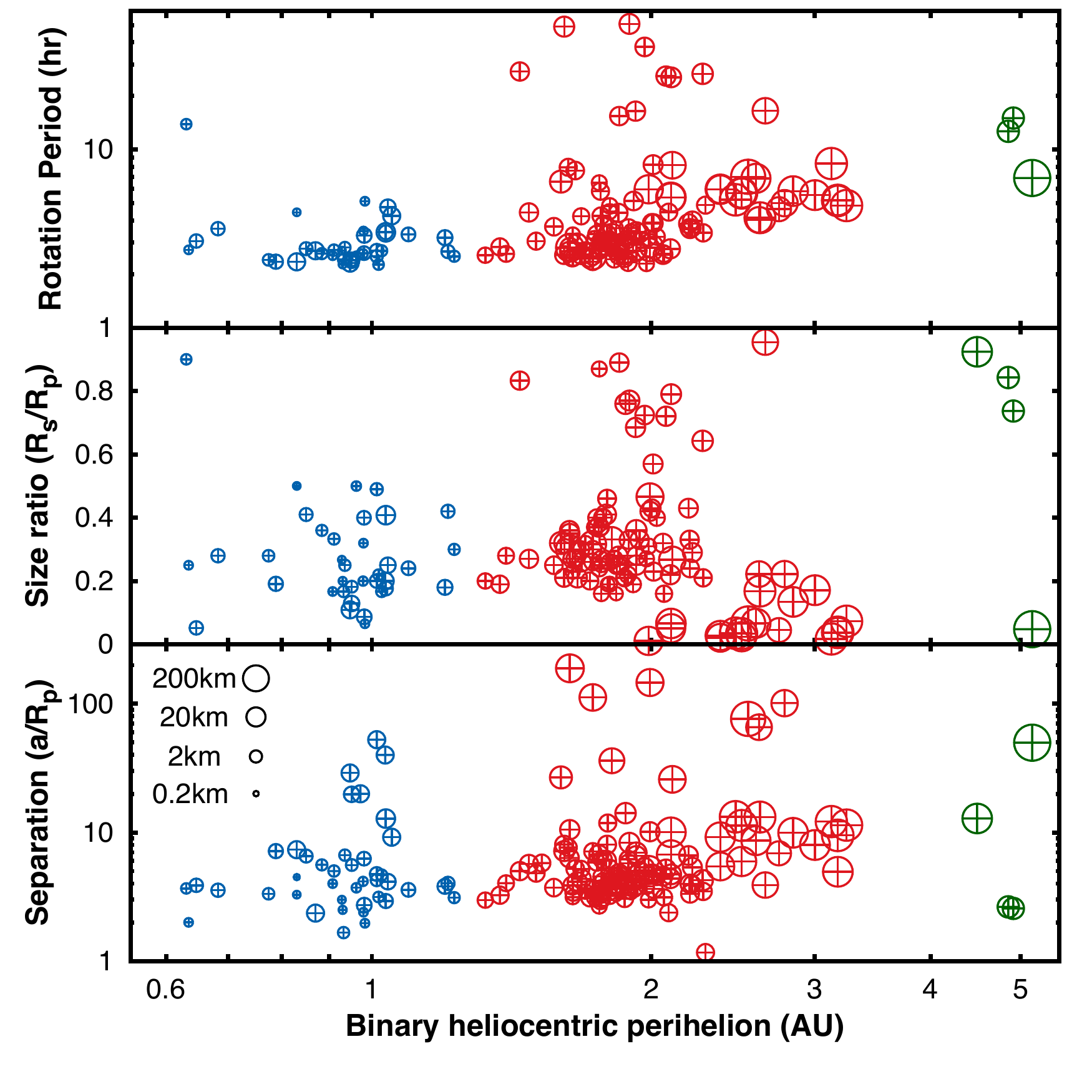}
\caption{\small The known population of binary asteroids. 
Three panels show (Bottom) the component separation in terms of the primary radius, (Middle) the size ratio of the two components, and (Top) the rotation period of the primary all plotted as a function of the system's heliocentric orbit's pericenter~\citep{Johnston:2014wm}.
The size of the symbol indicates the size of the primary body, with the scale being on the left side of the bottom panel.}
\label{fig1}
 \end{figure*}

The known binary systems among NEAs have primary component diameters
exclusively less than 10~km.  These small systems typically have
moderately-sized secondaries between 4--58\% the size of the primary
---corresponding to a mass ratio range of 6.4$\times$10$^{-5}$--
2.0$\times$10$^{-1}$ assuming equal densities, are on tight orbits
with typically 2.5-- 7.2~primary radii separation, and have a
fast-spinning primary with rotation period between 2.2--4.5~hours
---all are below twice the critical disruption spin period of 2.3
hours for a sphere of density 2 g cm$^{-3}$.  The data for these
systems is presented in Figure~\ref{fig1}.  When lightcurve surveys
probed similar sized asteroids in the Main Belt, they found systems
with the same characteristics existing at roughly the same proportion
of the population~\citep{Warner:2007uy}.

The known binary systems among MBAs have properties that vary with
their size.  The small population ($D < 15$ km; (4492) Debussy is the
largest) look similar to the various morphologies found among NEAs
including a few that appear similar to (69320) Hermes, while
asteroid satellites around large asteroids ($D > 25$ km; (243) Ida is
the smallest of this group) fall into other categories.  These larger asteroid
systems have by comparison much smaller satellites on much more
distant orbits.  While a number of the large asteroids with satellites
have rotation periods lower than the average asteroids of these
sizes~\citep[geometric means and 1-$\sigma$ deviations are $7.6 \pm
  0.4$ hours versus $12.2 \pm 0.5$
  hours;][]{Warner:2009ds,Pravec:2012fa}, their rotation periods are
all more than twice the critical disruption spin period of 2.3 hours
for a sphere of density 2 g cm$^{-3}$ with the exception of (22)
Kalliope~\citep{Pravec:2012fa,Johnston:2014wm}.

The techniques used to increase this database of known systems over
time are important as they define any biases of our knowledge of each
population.  Lightcurve techniques, which are important for finding
close companions around small asteroids, are strongly biased against
finding distant companions.  Meanwhile radar can discover satellites
widely separated from their primaries, but is ineffective for
observing distant Main Belt asteroids~\citep{Ostro:2002ut}.  Direct
high-resolution imaging is best for finding distant companions of
large Main Belt asteroids~\citep{Merline:2002vn}.  The size of the
known population of small binary asteroid systems has increased
substantially owing to the ready availability of small telescopes to
survey asteroid lightcurves and the increased frequency of radar
observations. Meanwhile many large Main Belt asteroids have been
surveyed with ground-based telescopes with far fewer recent
discoveries -- though new Adaptive Optics (AO) technologies may
uncover previously unseen satellites at previously studied
asteroids~\citep{Marchis:2014DPS}.

\bigskip
\noindent
\textbf{ 1.2 The YORP effect}
\bigskip

The largest shift in the understanding of binary formation and
evolution has come from studies of thermal forces that can affect a
single body or a binary system. The reflection and reemission of solar
radiation can produce a torque that changes the rotation rate and
obliquity of a small body. This effect is referred to as the YORP
effect, coined by~\citet{Rubincam:2000fg}, and evolved out of the work
of many researchers on similar
topics~\citep{Radzievskii:1952uo,Paddack:1969bu,Paddack:1975cq,OKeefe:1976vc}. We
only provide a brief summary here (see~\citet{Bottke:2006en} and the
Vokrouhlick{\'y} et al. chapter in this volume for a detailed
discussion of the effect).

The YORP effect has been directly detected for several asteroids
through observed rotation rate changes
~\citep{Taylor:2007kp,Lowry:2007by,Kaasalainen:2007hq,Durech:2008di,Durech:2008bn,Durech:2012bq}.
These rotation rate changes match the predicted magnitude of the
effect from theoretical
predictions~\citep{Rubincam:2000fg,BottkeJr:2002tu,Vokrouhlicky:2002cq,Capek:2004bl,Rozitis:2013kk}.
The YORP effect has a straightforward dependence on asteroid size
(timescales increase with $R^2$) and distance from the Sun (timescales
increase with $a^2$), but a complicated relationship with
shape~\citep{Nesvorny:2007gz,Scheeres:2007kv}.  This shape dependence
is characterized by a YORP coefficient, which measures the asymmetry
of the body averaged about a rotation state and a heliocentric orbit.
While instantaneous estimates of the YORP coefficient are available
from astronomical measurements of the radial accelerations of
asteroids, theoretical models of long-term averaged values are stymied
by a sensitive dependence on small scale
topography~\citep{Statler:2009fw,McMahon:2013ww,CottoFigueroa:2014tl}
and regolith properties~\citep{Rozitis:2012fq,Rozitis:2013da}.  NEAs
and MBAs with diameters below about 20 km are likely to be affected on
Solar System timescales~\citep{Bottke:2006en,Jacobson:2014hp}.
Kilometer-size NEAs can have rotation rate doubling timescales shorter
than their dynamical lifetime of $\sim$10~Myr and MBAs shorter than
their collision lifetime of
$\sim$100~Myr~\citep{Bottke:2006en,Jacobson:2014us}.

The distribution of spin rates observed for bodies smaller than 40 km
in size show excesses of very fast and slow
rotators~\citep{Pravec:2000dr,Warner:2007uy}, which is matched very
well by a spin distribution model including the YORP
effect~\citep{Pravec:2008cr,Rossi:2009kz,Marzari:2011dx} as suspected
in the {\it Asteroids III} chapter by~\citet{Pravec:2002ts}. Note that
the very large asteroid lightcurve survey of \citet{Masiero:2009et}
found a more Maxwellian distribution of spin rates among small
asteroids though it is not necessarily incompatible with YORP spin
evolution.  Among larger bodies, a subset of the Koronis asteroid
family was found to have aligned obliquity and clustered spin
rates~\citep{Slivan:2002iw}, which is due to the YORP effect driving
them into spin-orbit resonances~\citep{Vokrouhlicky:2003to}.  Amongst
asteroid families, whose spreading is controlled by the Yarkovsky
drift of family members, there are clear signatures of the YORP effect
changing the obliquity of smaller bodies and in turn changing their
Yarkovsky drift rates
~\citep{Vokrouhlicky:2006kf,Bottke:2006en,Bottke:2015jg}.

Morning and evening thermal differences across regolith blocks torque
the asteroid similarly in magnitude to the ``normal'' YORP effect.
Unlike the effect described above, this ``tangential'' YORP effect
does depend on the rotation rate, material properties of the regolith,
and size distribution of the
blocks~\citep{Golubov:2012kt,Golubov:2014hf}. Furthermore, the
tangential YORP effect has a prograde bias unlike the normal YORP
effect, which is unbiased.  This additional torque may explain the
difference between the predicted rotational deceleration of
Itokawa~\citep{Scheeres:2007bj,Durech:2008bn,Breiter:2009fw} and the
observed acceleration by the Japanese space mission
Hayabusa~\citep{Lowry:2014cb,Golubov:2014hf}.  Similarly, a preference
for spinning up may be necessary to explain the large fraction of
observed binary systems, $\sim$15\% of small asteroids, which are
presumed to be formed from rotational disruption caused by continued
YORP spinup (discussed in detail below).

As mentioned as early as the~\citet{Vokrouhlicky:2002cq} and
\citet{BottkeJr:2002tu} works on the YORP effect, this effect was a
very good candidate to rotationally disrupt rubble pile asteroids.  As
the catalog of known systems has grown, and the sub-populations of
binary systems became more defined, rotational disruption by the YORP
effect emerged as the primary candidate to be the dominant formation
mechanism.  Much of the recent research, and the discussion below, is
focused on the step(s) between when the YORP effect starts increasing
the angular momentum of an asteroid and when we observe the diverse
catalog of systems today.  Some sub-populations may emerge directly
from YORP-induced rotational disruption, while others seem to demand
further evolutionary forces.
  
\bigskip
\centerline{\textbf{ 2. Binary Sub-Populations}}
\bigskip

With over 100 systems spread between NEAs and MBAs, clear
sub-populations of binary systems have emerged.  \citet{Pravec:2007fw}
compiled the parameters for the catalog of known binary systems,
including a calculation of the total angular momentum of each system.
They used this data to create a classification system of the known
inner Solar System binary systems that is well suited for the topics
in this chapter, (most of these populations are clear in Figures
  1 and 2), and we update it below:

\begin{itemize}
\item{Group L:} Large asteroids (diameter: $D > 20$~km) with
  relatively very small satellites (secondary to primary diameter
  ratio: $D_2 / D_1 \lesssim 0.2$).  We identify 11 members.
\item{Group A:} Small asteroids ($D<20$~km) with relatively small
  satellites ($0.1 \gtrsim D_2 / D_1 \lesssim 0.6$) in tight mutual
  orbits (semimajor axis, $a$, less than 9 primary radii, $R_p$).  We
  identify 88 members.

\item Group B: Small asteroids ($D<20$~km) with relatively large
  satellites ($0.7 \gtrsim D_2 / D_1$) in tight mutual orbits ($a
  \lesssim$ 9 $R_p$).  We identify 9 members.
\item{Group W:} Small asteroids ($D < 20$~km) with relatively small
  satellites ($0.2 \gtrsim D_2 / D_1 \lesssim 0.7$) in wide mutual
  orbits ($a \gtrsim$ 9 $R_p$). We identify 9 members.
\item{Three outliers:} Two would-be Group L members (90
  Antiope and 617 Patroclus) have the Group B characteristic of
  similar sized components but are much larger than the other Group B
  members ($D \sim$ 87 and 101 km, respectively), and a would-be
    Group W member (4951 Iwamoto) has the Group B characteristic of
    similar sized components but a much wider mutual orbit ($a \sim 17
    R_p$).
\item{Split Pairs:} These are inferred systems due to dynamical models
  that {\it very} closely link their heliocentric orbits.  Therefore
  they are not actual binaries, but are rather inferred dynamical
  end-states.  Some pair members are binaries themselves.
\end{itemize}
The principal change in this classification scheme
from~\citet{Pravec:2007fw} is the size ---splitting ``large'' and
``small'' asteroids.  Previously, ``large'' was defined to be
asteroids with diameters larger than 90 km, but from more recent
binary asteroid observations and YORP theory, a more natural boundary
between ``large'' and ``small'' is 20
km~\citep{Pravec:2012fa,Jacobson:2014bi}.  The boundaries
    between the various defining characteristics appear robust in
    Figure 1, where adjustments on the order of 10\% lead to the
    creation of no or only a few new outliers.  The data in the
    paragraphs above come from Petr Pravec's binary
    catalogue~\citep{Pravec:2012fa}.

\bigskip
\noindent
\textbf{ 2.1 Large Systems---Group L}
\bigskip

These are distinct in the bottom two panels of Figure 1, and are
defined by having large primaries with $D>$20~km (large symbols), and
relatively smaller size ratios ($D_2 / D_1 \lesssim 0.2$).  These size
ratios range from 0.03 to 0.2 with the lowest mean of any group: 0.08
$\pm$ 0.06 (1-$\sigma$)~\citep{Pravec:2012fa}.  They are
  typically discovered with ground-based high-resolution imaging with
  (243) Ida a notable exception, whose satellite was discovered by the
  Galileo space mission~\citep{Belton:1994uf}. There are 11 known
  systems, with the largest being (87) Sylvia ($D\sim$ 256 km) with
  its satellites Romulus and Remus~\citep{Brown:2001tc,Marchis:2005fi}
  and the smallest being (243) Ida ($D\sim$ 32 km) with its satellite
  Dactyl~\citep{Belton:1994uf}.  None of the 8 asteroids larger than
  (87) Sylvia have been reported to have satellites but there are
  likely many satellites amongst the asteroids with sizes near (243)
  Ida and up to (87) Sylvia since severe biases limit detection in
  that population.

The rotation periods of these large asteroids range from 4.1--7.0 hr
with a geometric mean: 5.6 $\pm$ 0.8 (1-$\sigma$
hr)~\citep{Pravec:2012fa}. With the exception of (22) Kalliope, all of
the Group L members rotate at less than half the critical disruption
rate for a spherical body with a density of 2 g cm$^{-3}$ (2.3 hr
rotation period).  As discussed in~\citet{Descamps:2011}, the lower
than typical rotation periods~\citep[asteroids of similar sizes have
  12.2 $\pm$ 0.5 hr rotation periods;][]{Warner:2009ds} and the
elongated shapes of the primaries~\citep[e.g. the bean-shape of (87)
  Sylvia;][]{Marchis:2005fi} are suggestive of a violent disruption
process, with the reaccumulation of the parent body into high angular
momentum shape and spin configurations.  However, previous numerical
models of asteroid disruptions did not retain shape and spin
information of the re-accumulated remnants~\citep{Durda:2007im}, except in the case of Itokawa \citep{Michel:2013AA}, so
this has not been explicitly tested.

It is important to note that due to an increase of YORP
  timescales with surface area, YORP cannot play a role for these
  large systems.

\bigskip
\noindent
\textbf{ 2.2 Small Systems---Groups A, B and W}
\bigskip

Systems with small primary bodies ($D<20$~km) almost all fit in the
sub-population of Group A, and are found among NEAs, small MBAs and
Mars Crossers.  Members of Group A have diameters between 0.15 and 11
km~\citep{Pravec:2012fa}. Their mutual orbits are within 2.7 and 9.0
primary radii (mean statistics: 4.8 $\pm$ 1.3 $R_p$), and the
satellites are between 0.09--0.58 the size of the primary (mean
statistics: 0.3 $\pm$ 0.1).

Group A members have similar amounts of angular momentum relative to
their critical values (where critical is enough to breakup the
combined masses if in a single body). Typically this is due to the
rapidly rotating primary with a period between 2.2 and 4.4
hr~\citep{Margot:2002fe,Pravec:2006bc,Pravec:2012fa} ---always within
a factor of two of the critical disruption rate for a spherical body
with a density of 2 g cm$^{-3}$ (2.3 hr rotation period).  The
geometric mean of their rotation periods is 2.9 $\pm$ 0.8 (1-$\sigma$)
hr compared to 7.4 $\pm$ 0.3 hr for asteroids in the same size
range~\citep{Pravec:2012fa,Warner:2009ds}.

While the orbit period is never synchronous with the very fast primary
rotation (see Fig 1. and 2.), the synchronicity of the satellite
period with the orbit period divides Group A into two distinct
subgroups. Most belong to the synchronous satellite
subgroup~\citep[$\sim$66\%;][]{Pravec:2012fa}, although it is possible
that these satellites have chaotic rotation but appear nearly
synchronous~\citep{Naidu:2015gp}. The rest of the asteroids have
satellites that are asynchronous, and have rotation periods between
the primary rotation period and the orbit
period~\citep{Pravec:2012fa}.

There are very few well characterized mutual orbits---only seven
binary and two triple systems~\citep{Fang:2012fw}.  Among these there
is a trend between their measured orbital eccentricity and the
synchronicity of the satellites rotation and orbit periods (see Fig. 1
of~\citet{Fang:2012fw}), where synchronous satellites are generally
less eccentric.  This trend is consistent with limits on the
eccentricity determined by lightcurve studies~\citep{Pravec:2007fw}.
Some asynchronous satellites on eccentric orbits may be in chaotic
rotation as a result of torques on their elongated
shapes~\citep[e.g. 1991 VH;][]{Naidu:2015gp}.  This could explain the
high frequency ($\sim$33\%) of asynchronous satellites in light of
possibly rapid theoretical de-spinning timescales ---around
10$^2$--10$^8$ years~\citep[the tidal parameters are very
  uncertain;][]{Goldreich:2009ii,Jacobson:2011hp,Fang:2012fw} compared
to a $\sim$10$^7$ year dynamical lifetime in the NEA population or a
$\sim$10$^8$ year collisional lifetime in the MBA population.

The Group B members have nearly equal-sized components (mean
$D_2/D_1$ is 0.88 $\pm$0.09; they stand out in the middle panel of
Figure 1). There is a break amongst the small binary
population between Group A and Group B, suggesting an
alternate evolutionary route~\citep{Scheeres:2004fo,Jacobson:2011eq}.
These systems are in a doubly synchronous state with synchronized
rotation and orbital periods~\citep{Pravec:2007fw}.  Orbital periods
extend from 13.9 hr to 49.1 hr similar to the Group A
binaries~\citep[11.7 hr to 58.6 hr;][]{Pravec:2012fa}.  Nearly all of
this population is found in the Main Belt and discovered by lightcurve
observations with the exception of the NEA (69230) Hermes.  Hermes is
  decidely smaller ($D\sim$ 0.6 km; Margot et al. p.c.) than the next
  smallest confirmed Group B member~\citep[(7369) Gavrilin, $D\sim$
    4.6 km;][]{Pravec:2007fw}, but an unconfirmed member of Group B
  1994 CJ$_1$ is even smaller~\citep[$D<$ 0.15
    km;][]{Taylor:2014wa}. The mean size $D$ is 7 $\pm$ 3 km, and the
largest (4492) Debussy is 12.6 km~\citep{Pravec:2012fa}. Because these
systems are doubly synchronous, there are strong biases against
discovery; only mutual events reveal the presence of the satellite.
Furthermore, they are not typically separated enough to be detected by
high-resolution imaging (the widest is (854) Frostia with a component
separation of only 36.9~km) and too distant and small for radar
detection in the Main Belt. Thus they are possibly significantly
under-represented amongst known asteroid binaries.

Group W binary members have very large separations (between 9 and 116
$R_p$), and are typically detected by the Hubble Space Telescope or
adaptive optics observations ground-based telescope observatories. The
existence of some of this group in the Main Belt defies formation only
from planetary encounters ~\citep{Fang:2012go}.
Like Group A and B members, they are
  small (mean size $D$ is 4.8 $\pm$ 2.3 km)~\citep{Pravec:2012fa}, so
  radiative torques like the YORP effect are important.  Like Group A
  members, they are rapidly rotating \citep[geometric mean primary
    rotation period 3.3 $\pm$ 0.8 hr;][]{Polishook:2011if} and have
  moderate size ratios~\citep[mean size ratio $D_2/D_1$ is 0.4 $\pm$
    0.2;][]{Pravec:2012fa}.  They follow the pattern of the
  asynchronous subgroup of Group A and all of their satellites have
  rotation periods that are between that of the primary's rotation and
  the satellites orbital period.  The links between Group A and Group
  W are strong and~\citet{Jacobson:2014hp} developed an evolutionary
  pathway from the former to the latter.  Previously, Group W members
  have been suspected to be debris from catastrophic
  collisions~\citep[dubbed EEBs in ][]{Durda:2004en}, but further study
  consistently finds rapidly rotating
  primaries~\citep{Polishook:2011if}.

\begin{figure*}
 \plotone{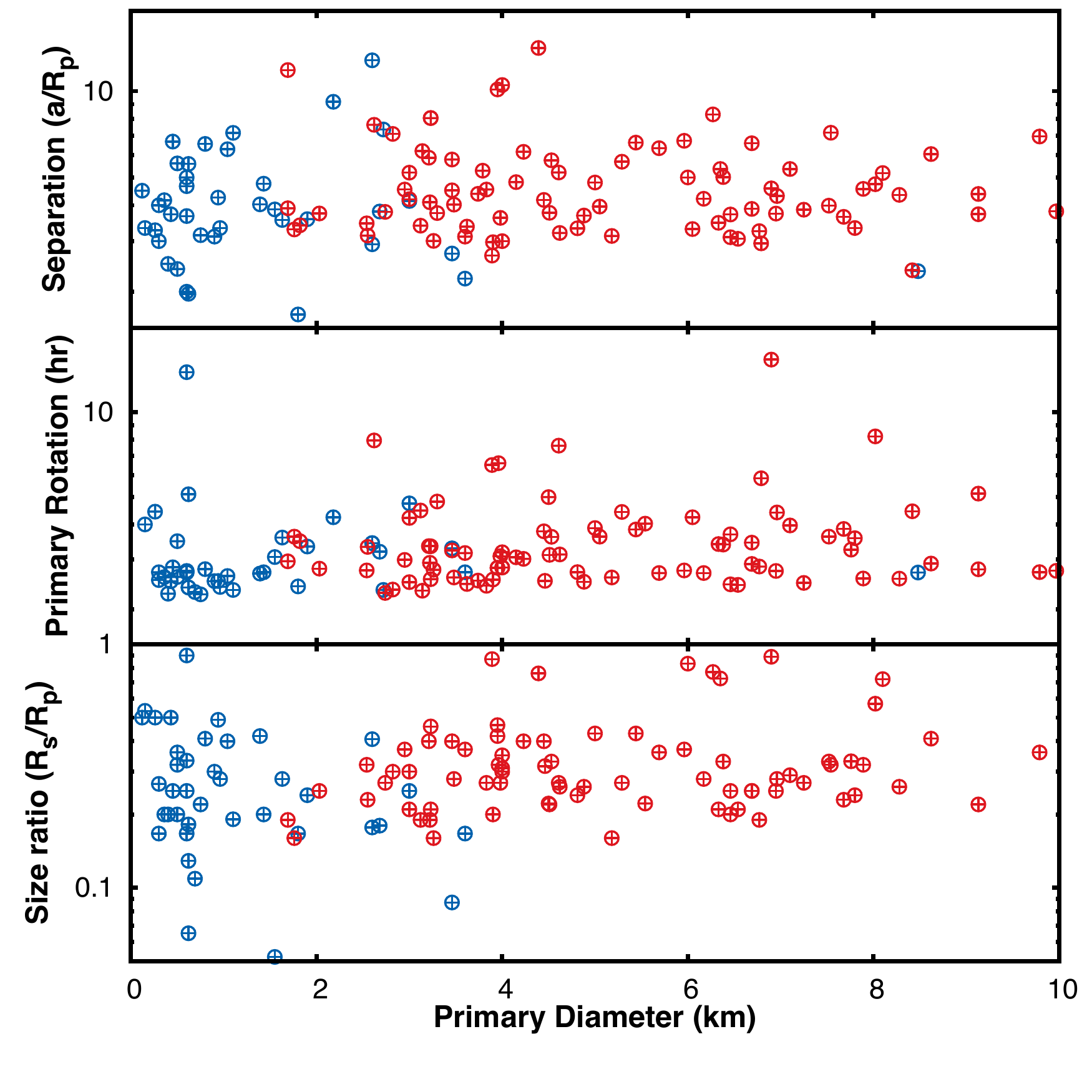}
 \caption{\small The population of small binary systems, showing their
   (Bottom) size ratio, (Middle) primary rotation period and (Top)
   component separation, all plotted as a function of their Primary
   Diameter~\citep{Pravec:2007fw,Johnston:2014wm} The NEAs are
   blue symbols and MBAs are red symbols.
\label{fig2}}
\end{figure*}

Many of these systems were discovered and characterized by lightcurve
observations, which produce rotation periods as well as some
information on shape from the amplitude of the
lightcurve~\citep{Pravec:2007fw}.  The lightcurve amplitude
principally constrains two of the axes of the body, the long $a$ and
intermediate $b$ axis (for principal axis rotation about its shortest
$c$ principal axis). It can constrain the other axis ratio if there
are multiple observing geometries.  The lightcurve amplitude can be
converted to determine the $a/b$ relationship, which roughly describes
the shape of the body's equatorial cross-section.  Group A, B and W
members have $a/b$ from 1.01 to 1.35 with an average value of 1.13
$\pm$ 0.07~\citep{Pravec:2012fa} ---nearly-circular equatorial
cross-sections. Meanwhile the satellites of Group A, B and W members
have $a/b$ from 1.06 to 2.5 with an average value of 1.44 $\pm$
0.24~\citep{Pravec:2012fa}. Thus satellites represent a much larger
variety of equatorial cross-sections.

Lastly, some primary members of Group A and W have a characteristic
spheroidal ``top'' shape due to a pronounced deviation from a sphere
along an equatorial ridge. This radar-derived shape was made famous by
1999 KW$_{4}$~\citep{Ostro:2006dq}, but has been found for many other
binary and single asteroids~((29075) 1950 DA \citep{Busch:2007fz};
2004 DC \citep{Taylor:2008ve}; 2008 EV$_5$ \citep{Busch:2011jr};
(101955) Bennu \citep{Nolan:2013gj}; (136617) 1994 CC
\citep{Brozovic:2011ib}; (153591) 2001 SN$_{263}$
\citep{Becker:2015il}). This ridge preserves a low $a/b$ ratio, i.e. a
circular equatorial cross-section, but due to the confluence of
rotation and shape this reduces the gravitational binding energy of
material on the
ridge~\citep{Ostro:2006dq,Busch:2011jr,Scheeres:2015hr}. At high
rotation rates, the entire mid-latitudes obtain high slopes and so
disturbed loose material would naturally move towards the potential
low at the equator; this material upon reaching the equator may move
off the surface entirely and enter into
orbit~\citep{Ostro:2006dq,Walsh:2008gk,Harris:2009ea}. This discovery
has driven studies of asteroid re-shaping focusing on the granular and
cohesive properties of the surface material and possible secondary
fragmentation and in-fall of orbital
material~\citep{Ostro:2006dq,Walsh:2008gk,Harris:2009ea,Holsapple:2010fv,Jacobson:2011eq,Scheeres:2015hr}.

\bigskip
\noindent
\textbf{ 2.4 Triples}
\bigskip

The first discovered triple in the Main Belt was (87) Sylvia, with two
small satellites orbiting its bean-shaped primary
body~\citep{Marchis:2005fi}.  More triples have since been found, with
(45) Eugenia~\citep{Marchis:2007tx}, (93)
Minerva~\citep{Marchis:2009ur} and (216) Kleopatra joining the
list~\citep{Marchis:2008tk}.  As discussed below, this is believed to
be a natural outcome of formation via asteroid collisions.

There are also a few confirmed and suspected small asteroid triple
systems (136617) 1994 CC, (153591) 2001 SN$_{263}$, 2002 CE$_{26}$,
(3749) Balam and (8306)
Shoko~\citep{Brozovic:2009ib,Nolan:2008ty,Shepard:2006fn,Marchis:2008tk,Pravec:2013ud}.
All have rapidly rotating primaries (2001 SN$_{263}$ is the lowest at
3.425~hours) and low size ratios between the smaller two members and
the primary (Balam has the largest measured satellite at 46.6\% its
size). For the two triple systems with known primary shapes ((136617)
1994 CC, (153591) 2001 SN$_{263}$), both have the typical
``top'' shape described above~\citep{Brozovic:2011ib,Becker:2009wj}.

Both Balam and Shoko are also members of split pairs and the other
members are 2009 BR$_{60}$ and 2011 SR$_{158}$,
respectively~\citep{Vokrouhlicky:2009ja,Pravec:2013ud}.  As explained
in the next subsection, split pairs have a dynamical age that is
interpreted as the rotational fission formation age.  Since it is
unlikely that the split pair member could have formed without
significantly affecting the triple system, it is possible that all
components were created at the same time from a single rotational
fission event~\citep{Jacobson:2011vk}.

\bigskip
\noindent
\textbf{ 2.5 Outliers}
\bigskip

Large double asteroids such as the MBA (90) Antiope and Trojan (617)
Patroclus appear unique in the inner Solar System.  These are too
large, with diameters greater than 100~km, to have gained angular
momentum from thermal effects and collision simulations do not
typically create such systems~\citep{Durda:2004en}.  They have very
large angular momentum content, owing to the similar-sized
components~\citep{Pravec:2007fw,Descamps:2007im,Michaiowski:2004ky}.
Antiope is notable as it is among the largest fragments in an asteroid
family owing to the exceptional size of Themis and its
family. Meanwhile, Patroclus is a Trojan, and Solar System formation
models suggest that many or all of them may have been implanted from
the primordial Kuiper Belt
region~\citep{Morbidelli:2005dr,Nesvorny:2013jw}.  Thus this system
may share a common origin with the systems found in the Kuiper
Belt~\citep[see][]{Noll:2008ku,Nesvorny:2010da}.

The other outlier, (4951) Iwamoto, has a much wider mutual orbit than other Group B members, but this may be explained by orbit expansion due to the BYORP effect as discussed below~\citep{Cuk:2007gr,Jacobson:2011vk}.

\bigskip
\noindent
\textbf{ 2.6 Split Pairs}
\bigskip

An important discovery related to the dynamics of binary systems is
the existence of individual asteroids that are not bound to each other
but instead show convincing signs of being split
pairs~\citep{Vokrouhlicky:2008dt,Vokrouhlicky:2009kt,Pravec:2009hv,Pravec:2010kt}.
These were found using dynamical studies similar to those that search
for families of asteroids, but here pairs were found to be closely
linked dynamically.

Follow-up observations have found convincing links in both size and
rotation of the pairs~\citep{Pravec:2010kt} and also the photometric
appearances~\citep{Moskovitz:2012ga,Duddy:2012gb}.  Their sizes and
rotation make a very strong case that the smaller member of the pair
was ejected during a rotational fission event, with the signature of
this in the slow rotation of the larger object as a function of the
size of the smaller object.  The latter work finds similar photometric
colors for the pairs, supporting the dynamical links between them.
The dynamical models suggest that some of these pairs separated less
than just $\sim$17~kyr ago, and hence the photometric colors have not
had time to evolve significantly due to space-weathering or other
effects~\citep{Vokrouhlicky:2009kt,Vokrouhlicky:2011ff}.

\bigskip
\centerline{\textbf{ 3. Formation}}
\bigskip

While the community and literature largely agree on the collisional
origin of large asteroid's satellites (Group L), there is continued
work on the details of how the small systems (Groups A, B, W) form and
evolve.  An important part of understanding the formation of the small
systems concerns both the properties and variety of outliers and the
possible complicated evolutionary paths for satellites or building
blocks of satellites once in orbit around a rubble pile asteroid.

\bigskip
\noindent
\textbf{ 3.1 Large Systems---Collision}
\bigskip

Collisions were proposed as a potentially important formation
mechanism even before the discovery of Ida's moon Dactyl in 1993.
Most works focused on ejecta from a collision becoming mutually bound,
becoming bound around the largest remnant, or rotational fission due
to a highly oblique or glancing
impact~\citep{Weidenschilling:1989vv,Merline:2002vn,Richardson:2006fo}.
Both~\citet{Weidenschilling:1989vv} and~\citet{Merline:2002vn} found
that complete disruption is a far more likely outcome than
collisionally-induced rotational fission, and there are no observed
systems that can be clearly attributed to this later process.

The study of the other collisional mechanisms first focused on
cratering events on the asteroid Ida, numerically tracking the
evolution of ejected debris in order to form its small satellite
Dactyl~\citep{Durda:1996fa,Doress:1997}.  Studies of asteroid impacts
gained a numerical boost by combining Smoothed Particle Hydrodynamics
models of asteroid fragmentation with $N$-body models of their
gravitational
re-accumulation~\citep{Michel:2001bw,Michel:2002jx,Michel:2003tb,Michel:2004jg,Durda:2004en,Durda:2007im}.
These models were more capable of modeling the physics of catastrophic
collisions and maintaining high-resolution models of the fragments
long-term gravitational interactions and re-accumulation. They found
that the formation of satellites is a natural outcome in an asteroid
disruption.

\citet{Durda:2004en} further explored the different types of systems
formed during a collision.  This large suite of 161 impact simulations
studied 100~km basalt targets being impacted by impactors of various
sizes hitting at a range of velocities and angles.  In their suite of
collision and re-accumulation simulations they observed, and named,
the two previously proposed types of systems: Escaping Ejecta Binary
systems (EEBs) and SMAshed Target Satellites (SMATS). The SMATs,
generally featured small satellite(s) orbiting the re-accumulated
target body.  The known satellites around large ($D$ $>$ 10~km) Main
Belt asteroids share similar properties---extreme size ratio between
primary and secondary and large orbital separation (where the
  orbital separations are too large to be explained by tides: see
  Section 4.2).  They predicted a formation rate that should roughly
produce the observed number of satellites detected around very large
asteroids ($D$ $>$ 140~km) accounting for their production due to
collisions, satellites destroyed by collisions and the very early
clearing of the asteroid belt.

Meanwhile they proposed that some small Main Belt systems featuring
two small components of roughly similar size on distant orbits are
possibly EEBs.  Their examples were (3749) Balam and (1509)
Esclangona, and while at the time both were interesting candidates,
(3749) Balam has been discovered to have a third component and a split
pair and (1509) Esclangona has been found to have a very rapid
rotation period similar to that found among many of the binary systems
formed by the YORP effect~\citep{Warner:2010vb}.  The best remaining
candidates are (317) Roxane {because of its} slow primary
rotation~\citep[8.2
  hr;][]{Harris:1992ki,Polishook:2011if,Jacobson:2014hp} and (1717)
Arlon because of its slow primary rotation~\citep[5.1
  hr;][]{Cooney:2006ur} and high size
ratio~\citep[$>$0.22;][]{Cooney:2006ur,Jacobson:2014hp}.  The lack of
EEBs in the known catalog is curious, as the simulations
of~\citet{Durda:2004en} formed hundreds of systems immediately after a
collision, although the stability of these binaries was not thoroughly
examined.  This is an important avenue for future work especially
given the importance of spin-orbit coupling for binaries after
rotational fission~\citep{Jacobson:2011eq}.  Tens of asteroid families
are known (see Chapter by Nesvorn{\'y} et al.) and there is evidence
for even very recent impacts throughout the Solar
System~\citep{Nesvorny:2002vc,Vokrouhlicky:2009kt}.  However, small
and wide binary systems are difficult to find and small components are
more susceptible to collisional grinding (see Chapter by Bottke et
al.), which may explain the lack of discoveries of this type of
system. Meanwhile the known systems need substantial characterization
(rotation periods etc.) to try to distinguish between possible EEBs
and end-states of YORP/BYORP evolution processes (see below).

All of the numerical work to understand formation of satellites during
collisions have found triple and multiple systems in their
simulations.  \citet{Durda:2004en} reported temporary multiple
systems, and Leinhardt \& Richardson (2005) re-analysis of a single
simulation found 10\% triples and 3\% quadruple systems that lasted
the length of the simulations (days).  While triples have now been
found among some large systems, longer-term dynamical simulations of
their formation and evolution following large impacts would be needed
to quantify the match between observations and models.

Catastrophic impact modeling has generally relied on very similar
collision scenarios (impact speeds and angles etc.) to model both the
formation of satellites and asteroid
families~\citep{Michel:2001bw,Michel:2003tb,Durda:2004en,Durda:2007im}.
While asteroid families are strictly correlated with collisions,
it does not mean that the presence of a family demands satellites, as
not every collision forms satellites, and small satellites themselves
are susceptible to collisional evolution/destruction on timescales
shorter than the age of many observed asteroid
families~\citep{Durda:2004en}.

\bigskip
\noindent
\textbf{ 3.1 Small Systems---Rotational Disruption}
\bigskip

Even before the discovery of small binary systems, the doublet craters
found on the terrestrial planets~\citep{Melosh:1991cp,Bottke:1996ek}
and crater chains on the Moon~\citep[see][]{Richardson:1998bq}
suggested that there were mechanisms to disrupt small asteroids.  The
demonstration provided by Comet Shoemaker-Levy 9 and its tidal
disruption at Jupiter further instigated models of ``rubble pile''
interiors and their tidal disruptions while encountering planetary
bodies~\citep{Asphaug:1996jh,Richardson:1998bq}

\citet{Bottke:1996ek} suggested that searches for asteroid satellites
``place emphasis on kilometer-sized Earth-crossering asteroids with
short-rotation periods'', and lightcurve surveys found many
interesting targets in this sample.  Observations of multi-frequency
lightcurves and possible eclipse/occultation events became common, and
gave very strong indications of possible
satellites~\citep{Pravec:1997hf,Pravec:1998ke,Mottola:2000ct,Pravec:2000gw}.
The radar imaging of near-Earth asteroid 2000 DP$_{107}$ confirmed
that the lightcurve observations were detecting actual
satellites~\citep{Margot:2002fe}.  Combining all possible
detections~\citet{Margot:2002fe} suggested that up to 16\% of the
population were binaries and that rotational disruption was a primary
culprit~\citep{Pravec:1999wt,Margot:2002fe,Pravec:2007fw}.

Rubble pile asteroids encountering the Earth were studied with a
granular dynamics code by~\citet{Richardson:1998bq} and again
by~\citet{Walsh:2006gk}.  While both groups found that binaries are at
least initially formed following some disruptive tidal
event,~\citet{Walsh:2006gk} found that the primary bodies were
typically elongated, the secondaries were on very eccentric orbits
and the primary rotated with period around 3.5--6 hours, rather than
the near-critical 2-4~hr periods.~\citet{Walsh:2008gx} took the
resulting simulation outcomes and built a Monte Carlo model including
the expected time between planetary encounters, expected encounter
outcomes, nominal tidal evolution of orbits and primary spin, and
observed asteroid shape and spin characteristics.  They found that the
produced systems are not expected to survive very long, owing to the
large semimajor axes and high eccentricities.  These works, and the
discovery of small binary systems in the Main
Belt~\citep{Warner:2007uy} where there is no planetary body to tidally
disrupt an asteroid, strongly suggested that tidal disruption is not a
primary mechanism.  Tidal disruption of NEAs could still account for a
small subset of the population, though it is not clear if the
elongated primaries and eccentric secondaries could survive long
enough to be observed~\citep{Walsh:2008gx}.

A more ubiquitous method for rotational disruption of small asteroids
is the YORP effect.  \citet{Rubincam:2000fg} proposed that the YORP
effect could spin cm-sized objects so fast that they would eventually
burst.  \citet{Vokrouhlicky:2002cq} pointed out that this effect will
likely drive asteroids to 0$^\circ$/180$^\circ$ obliquity end-states
and then in many cases of continued spin-up could drive them to
``rotational fission''.  YORP was connected directly with binary
formation in {\it Asteroids III}~\citep{BottkeJr:2002tu}, where it was
proposed as a possible means for forming small binary asteroids and
inducing re-shaping.

\begin{figure*}
 \plotone{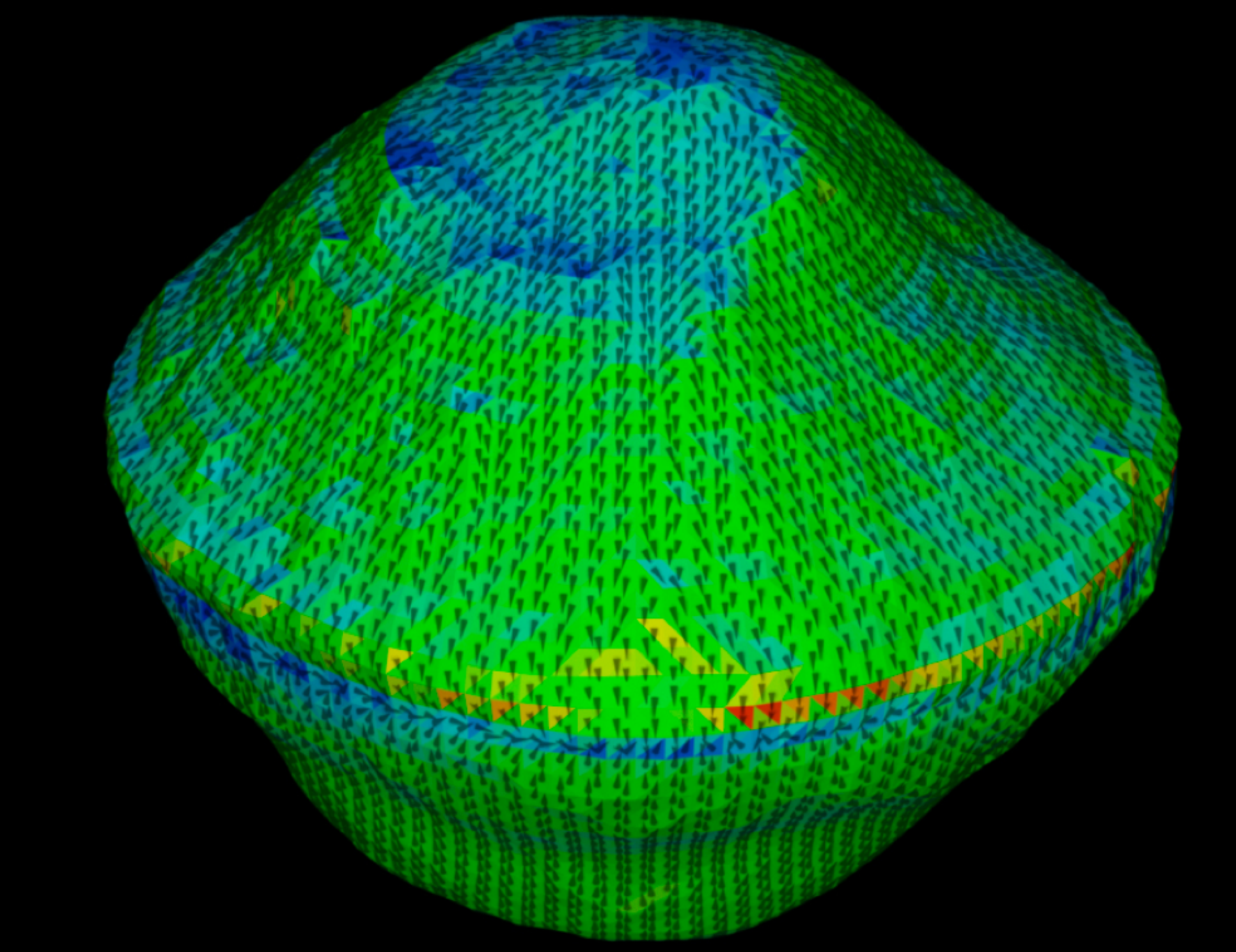}
\caption{The radar-derived shape model of 1999 KW$_4$. The coloring
  indicates local gravitational slopes with green's being near 30-40
  and the blue colors at the equator and poles near zero
 ~\citep{Ostro:2006dq,Scheeres:2006jc}.}
\label{fig:KW4}
\end{figure*}

\citet{Ostro:2006dq} observed NEA 1999 KW$_{4}$ with radar and
produced an incredibly detailed shape model of the primary
while~\citet{Scheeres:2006jc} analysed the dynamics of the
system~\citep[see][]{Fahnestock:2008gw}.  This system was similar to
previously discovered NEA systems---it featured a rapidly rotating
primary (essentially at critical rotation rate) and a small secondary
on a close orbit just beyond its Roche Limit.  However, owing to the
exceptional resolution of these radar observations, the derived shape
model was found to have a bulging equatorial ridge (see Fig
\ref{fig:KW4}).

As part of the dynamical analysis of the
system~\citet{Scheeres:2006jc} hypothesized that the system disrupted
and shed mass due to tidal torques from a planetary flyby or the YORP
effect.  The primary would have evolved to build the ridge and reach
the very rapid rotation rate due to the in-fall of material that was
not accreted/incorporated in the satellite.

Starting with the~\citet{Scheeres:2006jc} work on the state of the
1999 KW$_4$ system, and their suggestion that the equatorial ridge
could have been formed by the in-fall of material---this started one
of two tracks of thought about the ridge and its formation which were
followed up in a number of
works~\citep{Scheeres:2007kv,Jacobson:2011eq}.  These works posited
that the mass-loss was a more singular catastrophic event---a fission
event---and that later processing of this lost mass accounts for the
equatorial ridge and other widely observed system properties.  At the
other end of the discussion,~\citet{Walsh:2008gk} modeled YORP-spinup
of rubble piles made of thousands of constituent particles and posited
that the equatorial ridge was caused by re-shaping of the primary
rubble-pile asteroid as a result of spin-up and consequent mass-loss.
In this model the satellite was slowly built in orbit by repeated
mass-shedding events.

These models in some ways were working from opposite ends of a
spectrum of model resolution and techniques.  The
\citet{Scheeres:2007kv} work focused on the rigid body dynamics of
separated contact binaries, and~\citet{Jacobson:2011eq} extended this
to consider what might happen if the ejected fragment itself was
allowed to fragment once in orbit, which is critical to prevent rapid
ejection of the fragment, and also through the fall-in of some of the
material could explain the ubiquitous top-shape primaries.  Meanwhile,
~\citet{Walsh:2008gk,Walsh:2012jt} started with model
asteroids constructed of thousands of individual solid spherical
particles interacting through their gravity and through mutual
collisions.  While the gravity and collisions of the particles are
efficiently modeled throughout the simulations, the timescales for
spin-up were necessarily shortened for computational reasons and the
structure of the body consisted of different, but very simple, size
distributions of spherical particles.  A valuable test of these
  different ideas may occur when NASA's OSIRIS-REx space mission
  reaches asteroid Bennu, which shows signs of having an equatorial
  ridge~\citep{Nolan:2013gj,Lauretta:2014hi}.

All scenarios suffer confusion from new studies of the sensitivity of
the YORP effect to very small changes of an asteroid's shape. Model
asteroids were generated, inclusive of small features such as boulders
and small craters, and when YORP evolutions were calculated it was
found that very small changes on the surface of a small body can
dramatically change its YORP
behavior~\citep{Statler:2009fw,CottoFigueroa:2015}. The changes
could be so dramatic that nearly {\it any} re-shaping of a body during
its YORP spin-up could essentially result in a completely different
YORP-state. Essentially each shape-change, no matter the scale,
results in a coin-flip outcome to determine if the body continues
spinning-up or reverses and spins-down.

The population of asteroids with secondaries is $\sim$15\%
\citep{Pravec:2007fw}, and so the rotational disruption mechanism
appears to be quite efficient. If each movement on the surface of an
asteroid results in a coin-flip to determine spin-up or spin-down,
then seemingly bodies would never spin up enough to rotationally
disrupt.  This perplexing issue may demand some underlying tendency
for small bodies to ``spin-up'' by the YORP effect. It is
  possible that the YORP effect could actually induce preferential
  spin-up for even a symmetrically shaped asteroid, following the
  ``Tangential YORP effect'', which may play a big role in
  understanding these issues~\citep{Golubov:2012kt,Golubov:2014hf}.

\bigskip
\noindent
\textbf{ 3.3 Split Pairs}
\bigskip

The rotational fission hypothesis states that at a critical spin rate
an asteroid's components enter into orbit about each other from a
state of resting on each other~\citep{Jacobson:2011eq}. The spin
energy of the asteroid at this critical spin and any released binding
energy is the free energy available to disrupt the asteroid
system. Therefore, there is a direct energy and angular momentum
relationship between the spin states of the newly formed components
and the mutual orbit. From these considerations and some simple
assumptions, this theory predicts a relationship between the sizes of
the two components and the rotation rate of the larger
component. Observations of split pairs directly confirm this
theoretical prediction~\citep{Pravec:2010kt}. This is powerful
evidence that the rotational fission hypothesis is correct for split
pairs.

Further observations of split pairs confirm that each member is a good
spectroscopic match to the
other~\citep{Moskovitz:2012ga,Duddy:2012gb,Polishook:2014ub}. Interesting
observations that there is no significant longitudinal spectroscopic
variations and that the spin axes between members are identical are
interesting twists that future theory must account
for~\citep{Polishook:2014ub,Polishook:2014ul}.

\bigskip
\centerline{\textbf{ 4. Evolution of Binary Systems}}
\bigskip

There are a number of of different binary evolution
  mechanisms. Classical solid body tides are long studied and binary
asteroids provide useful test cases. Meanwhile, thermal effects can
affect a single body in the system or the pair of bodies. A single
body having its spin state changed by a thermal effect can possibly
re-shape due to its angular momentum increase. Binaries on
near-Earth orbits can encounter the terrestrial planets, which can
destabilize or otherwise alter a system's mutual orbit, while also
distorting or disrupting either component. Finally, an impact can
destroy a satellite, remove it from a system or simply perturb its
orbit.

The small number of known large systems are unaffected by many of
these evolutionary mechanisms: their satellites are typically too
distant for tides and their sizes too large for thermal effects, and
there are no planets in the Main Belt to perturb them. Meanwhile, the
known small system's may be affected by multiple effects
simultaneously in ways that are difficult to disentangle.  Therefore,
the primary set of data used to understand evolutionary effects are
the large number of small systems---Groups A, B and W. The majority of
all systems (A) look quite similar---they have rapidly rotating
primaries, secondaries just beyond the nominal Roche limit at
$\sim$2.5~$a/R_{p}$, where $a$ is the semimajor axis and $R_{p}$ is
the primary radius, and are between 9-58\% the size of the primary
(typically less than 2-3\% of the primary mass). The outliers are a
minority, but they and the split pairs point to important evolutionary
end-states of the small systems.

\bigskip
\noindent
\textbf{ 4.1 Binary-YORP}
\bigskip

The theory of binary YORP (BYORP) is a direct extension of the
Yarkovsky and YORP; instead of modifying the spin state of an
asteroid, the BYORP effect modifies the mutual orbit of a double
asteroid system in a spin-orbit resonance, typically the synchronous
1:1 spin-orbit resonance~\citep{Cuk:2005hb}.  Similar to YORP, the
back reaction force from the photon causes a torque, but here the
lever arm connects the center of mass of the binary system to the
emitting surface element. The back reaction torques the satellite
about this mass center changing the mutual orbit.  Unlike the YORP
effect, the relative position and orientation of the emitting surface
element can change the mutual orbit's semimajor axis $a$ with respect
to the center of mass of the binary system (unlike a rigidly rotating
asteroid in the case of the YORP effect), so only binary members that
occupy a spin-orbit resonance have non-zero cumulative BYORP effects;
the torques on all binary members outside of spin-orbit resonances
cancel out over time.

\citet{Cuk:2005hb} recognized that this effect could be significant
for small asteroids found throughout the binary asteroid population.
Most discovered binaries in the near-Earth and Main Belt populations
have small (Radius~$<10$ km) secondaries, which are tidally locked in a
synchronous spin-orbit
resonance~\citep{Richardson:2006fo,Pravec:2006bc}.  From these
characteristics and shape estimates, simple estimates scaled from the
YORP effect concluded that the BYORP effect is able to significantly
modify an orbit in as little as $\sim10^5$
years~\citep{Cuk:2005hb,Cuk:2007gr,Goldreich:2009ii}.  Secular
averaging theory has agreed with these short timescales
estimates~\citep{McMahon:2010by,McMahon:2010jy,Steinberg:2011jc}.

Assuming the smaller secondary is synchronously rotating and expanding
the solution to only first order in eccentricity, the secular
evolution of the mutual orbit's semi-major axis $a$, measured in
primary radii $R_p$, and eccentricity $e$ are~\citep[equations 93 and
  94 from][with re-defined variables]{McMahon:2010jy}:
\begin{subequations}
\begin{align}
\frac{da}{dt} = & \frac{3 H_\odot B_s a^{3/2} \sqrt{1+q}}{2 \pi \rho \omega_d R_p^2 q^{1/3}} \\
\frac{de}{dt} = &-\frac{3 H_\odot B_s a^{1/2} e \sqrt{1+q}}{8 \pi \rho \omega_d R_p^2 q^{1/3}} = -\frac{e}{4 a} \frac{da}{dt}
\label{eqn:byorp}
\end{align}
\end{subequations}
where $q$ is the mass ratio between the secondary and the primary, $\rho$ is the density of both asteroids assumed to be the same since they are likely to be of common origin, $\omega_d = \sqrt{4 \pi \rho G/3}$ is the critical rotational disruption rate for a sphere of density $\rho$, $G$ is the gravitational constant, $H_\odot = F_\odot / \left( a_\odot^2 \sqrt{1-e_\odot^2} \right)$ is a heliocentric orbit factor, $F_\odot$ is the solar radiation constant, and $a_\odot$ and $e_\odot$ are the heliocentric semi-major axis and eccentricity of the binary asteroid system. Lastly, $B_s$ is the BYORP coefficient of the secondary. 

As defined here, $B_s$ does not depend on the size of the secondary,
only its shape relative to its
orientation~\citep[see][]{McMahon:2010jy}. The BYORP coefficient can
be positive corresponding to outward expansion of the mutual orbit or
negative corresponding to inward shrinking.  (66391) 1999 KW$_4$ has
the only existing detailed secondary shape model~\citep{Ostro:2006dq},
and it has an estimated magnitude of $|B_s| \sim
0.04$~\citep{McMahon:2010jy,McMahon:2012ty}.  Estimates of $B_s$, from
other asteroid shape models and gaussian ellipsoids suggest that the
BYORP coefficients are typically $|B_s| <
0.05$~\citep{McMahon:2012ti}. Scaling the (66391) 1999 KW$_4$
  estimate to other binary asteroid systems,~\citet{Pravec:2010tc}
  calculated mutual orbit evolution predictions for seven observable
  binaries: (7088) Ishtar, (65803) Didymos, (66063) 1998 RO$_1$,
  (88710) 2001 SL$_9$, (137170) 1999 HF$_1$, (175706) 1996 FG$_3$ and
  (185851) 2000 DP$_{107}$.  First results regarding (175706) 1996
  FG$_3$ have been reported in~\citet{Scheirich:2015ez} and are
  discussed below.  Close observations of these candidates over the
  next few years will test the nascent BYORP theories.

As noticed initially by~\citet{Cuk:2005hb}, outward BYORP expansion of
the mutual orbit damps the eccentricity.  This potentially provides a
disruption pathway for binary asteroids.  Their orbit can expand until
the semi-major axis reaches their Hill radii since outward expansion
is a runaway process.  If so, then these binaries would become unbound
by three body interactions with the Sun and create asteroid pairs.
Unlike most observed asteroid pairs, these would not follow the
rotation-size ratio relationship set by immediate disruption after
rotational fission~\citep{Scheeres:2004fo,Pravec:2010kt}. No such
pairs have yet been identified, however the expected ratio between
pairs formed from fission to those formed from BYORP expansion is
high~\citep{Jacobson:2011eq}.

BYORP expansion of the orbit of the secondary will only continue if
the rotation of the secondary remains synchronous with its orbital
period. However, a numerical experiment by~\citet{Cuk:2010im} found
that the eccentricity may actually increase.  Eccentricity growth
induces chaotic rotation which is then halted by the BYORP effect, and
if the orientation of the secondary is reversed, then the mutual orbit
will contract. \citet{Cuk:2010im} rule out the role of the evection
resonance for responsibility of this eccentricity increase and
attributes it to spin-orbit coupling. This disagrees with evolution
resulting from the force decomposition and averaged
equations~\citep{Cuk:2005hb,Goldreich:2009ii,McMahon:2010by,Steinberg:2011jc}.
Future work directly comparing long-term evolution of
a~\citet{Cuk:2010im} type model and the secular evolution equations is
needed to determine resolutely the consequences of outward BYORP
evolution on eccentricity.

Using the secular evolution equations and including mutual
tides, ~\citet{Jacobson:2014hp} found that outward expansion can be
interrupted by an adiabatic invariance between the mutual semi-major
axis and libration state of the secondary.  As the mutual orbit
expands, a small libration can grow until the rotation of the
secondary de-synchronizes and begins to circulate.  This has been
proposed as the mechanism to explain the small known population of
wide binary asteroid systems that are found among NEAs and in the Main
Belt, as this process can leave secondaries stranded so far from the
primary to make tidal synchronization timescales very
long~\citep{Jacobson:2014hp}.  As observations continue to be made of
wide and possibly expanding binaries such as (185851) 2000 DP$_{107}$,
the theories regarding expansion due to the BYORP effect will continue
to be tested.

The BYORP effect can also shrink orbits and simultaneously increase
eccentricity~\citep{Cuk:2005hb,Goldreich:2009ii,McMahon:2010by,Steinberg:2011jc}.
This will be discussed after describing tides, which are important
when considering very tight binary asteroid systems.

\bigskip
\noindent
\textbf{ 4.2 Tides}
\bigskip

The evolutionary consequences of mutual body tides have been
considered for the evolution of asteroids since the discovery of the
first asteroid satellite, Dactyl about (243)
Ida~\citep{Petit:1997hb,Hurford:2000bp}.  These body tides are the
result of the asteroid's mass distribution chasing an ever-changing
equilibrium figure determined by the asteroid's spin state and the
gravitational potentials of both binary members.  Since the relaxation
towards this figure is dissipative, energy is lost in the form of heat
and removed from the rotation state of the asteroid.  The difference
in potential between the delayed figure and the theoretical
equilibrium figure is referred to as the tidal bulge.  This tidal
bulge torques the mutual orbit ensuring conservation of angular
momentum within the asteroid system.  Unlike lunar tides on Earth,
where most of the energy is dissipated at the ocean-seabed interface
and in the deep ocean
itself~\citep{Taylor:1920dp,Jeffreys:1921ii,Egbert:2000wm}, the mutual
tides between asteroids do not dissipate energy along an interface or
in a fluid layer but throughout the solid body.  However, new results
indicate that tidal dissipation in rubble piles may be much
higher~\citep{Scheirich:2015ez} than previously
expected~\citep{Goldreich:2009ii}, so where and how tidal energy is
dissipated must be examined much more thoroughly in the future.

Under most proposed formation circumstances and observed in nearly all
small systems ---except those with synchronous rotations, asteroids
rotate at rates greater than their mutual orbit mean
motion~\citep{Pravec:2006bc}.  In this case, the tidal bulge lags
behind the line connecting the mass centers of the binary members, and
the binary's primary is rotationally decelerated.  In this case, the mutual
orbit expands, similar to the Earth-Moon system.  Alternatively, the
tidal bulge precedes the line connecting the two asteroids, and the
binary's primary is rotationally accelerated.  In this case, the mutual orbit
shrinks, similar to the Mars-Phobos system.  No observed binary
asteroids currently occupy this state.  A third tidal state also
exists, librational tides can oscillate through tidally locked binary
members. This tide is responsible for removing libration from
synchronous satellites. The tidal bulge oscillates from the trailing
to leading hemisphere as the secondary librates, so the torque on the
orbit cancels out and the orbit does not evolve.

Formally deriving an explicit set of equations to describe these
torques has been a focus of research for over a
century~\citep{Darwin:1879vf}.  Historically, most theoretical
descriptions of tides have fallen into two camps split by their
assumptions regarding the relationship between the tidal bulge and the
line connecting the mass centers of the two asteroids: (1) some assume
a constant lag
angle~\citep{Goldreich:1963wa,Kaula:1964ex,MacDonald:1964gc,Goldreich:1966ky,Taylor:2010dk}
and (2) some assume a constant lag
time~\citep{Singer:1968ul,Mignard:1979kb,Mignard:1980bd,Hut:1980vi,Hut:1981vc}.
Neither relationship is expected to accurately reflect potential
asteroid
rheology~\citep{Efroimsky:2009jj,Greenberg:2009kt,Goldreich:2009ii,Jacobson:2011hp,FerrazMello:2013dq}.
Although the constant lag angle is believed to better represent
circulating tides through solid bodies,~\citet{Greenberg:2009kt}
describes its shortcomings in vivid starkness.  For the sake of this
review, we will consider using the theory only for nearly circular
orbits and will be careful to state when we feel that this theory may
not be adequate.  If systems have a non-negligible eccentricity, the
tidal de-spinning calculated by the first order theory will be a lower
bound, but the effects on the orbital evolution, particularly the
eccentricity, are more difficult to determine.  For instance,
the theories in~\citet{Goldreich:1963wa} and~\citet{Hut:1981vc} give
different predictions regarding the orbital evolution of asynchronous
asteroids depending on orbital parameters.

Constant lag angle tidal theory assumes that the tidal bulge raised by
an orbiting companion lags the line connecting the bodies' centers by
a constant angle $\epsilon$, which is related to a tidal dissipation
number $Q$ via: $Q = 1 / 2 \epsilon$.  The tidal dissipation number
quantifies the amount of energy dissipated each tidal frequency cycle
over the maximum energy stored in the tidal distortion~\citep[see the
  following for further
  discussion:][]{Goldreich:1966ky,Greenberg:2009kt,Efroimsky:2009jj}.
This theory is appropriate for determining the tidal torque on a
circulating body, but as the body approaches synchronization and when
the body is librating, this theory likely overestimates the actual
tidal torque.  The circulating tidal torque on a spherical asteroid
with radius $R$ from a perturbing binary member with a mass ratio of
$q$ is:
\begin{equation}
\Gamma_C = \frac{2 \pi k_2 \omega_d^2 \rho R^5 q^2}{Q a^6} \left( \frac{\omega-n}{|\omega-n|} \right)
\label{eqn:tidaltorque}
\end{equation}
where the semi-major axis $a$ is measured in asteroid radii
$R$, $\left( \omega-n / |\omega-n| \right)$ indicates the direction of
the torque given the spin rate of the asteroid $\omega$ and the mean
motion of the mutual orbit $n$, and $k_2$ is the second order Love
number of the asteroid.  The potential Love number $k$ quantifies the
additional gravitational potential produced by the tidal bulge over
the perturbing potential.  In other words, it captures how much the
tidal bulge responds to the deforming potential.  We are currently
considering only the lowest order relevant surface harmonic, namely
the second~\citep[see the following for a further discussion of the
  perturbing potential and its expansion,][]{FerrazMello:2008fz}.  A
perfectly rigid asteroid would have a tidal Love number of zero,
whereas a inviscid fluid would have a Love number of $3/2$ according
to its definition~\citep{Goldreich:2009ii}.

This tidal torque is most applicable to the primary, which is often
rapidly rotating compared to the mean motion of the mutual
orbit~\citep{Pravec:2006bc,Richardson:2006fo}.  In the most common
case, the secondary is tidally locked and so does not contribute to
the evolution of the semi-major axis of the mutual orbit.  Given the
torque above, the semi-major axis $a$, measured in primary radii
$R_p$, and the primary spin rate $\omega_p$ evolution
are~\citep{Goldreich:2009ii}:
\begin{subequations}
\begin{align}
\frac{da}{dt} = & \frac{3 k_{2,p} \omega_d q \sqrt{1+q}}{Q_p a^{11/2}} \\
\frac{d \omega_p}{dt} = & - \frac{15 k_{2,p} \omega_d^2 q^2}{4 Q_p a^6}
\end{align}
\end{subequations}
where $k_{2,p}$ and $Q_p$ are the tidal Love and dissipation numbers for the primary~\citep[for higher order expansions, see][]{Taylor:2010dk}. 

The ratio of tidal de-spinning timescales for the primary and secondary are:
\begin{equation}
\frac{\tau_s}{\tau_p} = \frac{k_{2,p} Q_s}{k_{2,s} Q_p} q^2
\end{equation}
where $k_{2,s}$ and $Q_s$ are the tidal Love and dissipation numbers
for the secondary. Since the mass ratio is often of order $q \sim
0.01$--$0.1$, immediately it is clear that the secondary tidally locks
first. It is possible that the ratio of tidal parameters could
counteract this, however both the tidal parameters derived from a
modified continuum tidal theory~\citep{Goldreich:2009ii} and the
observed parameters from a hypothetical tidal-BYORP
equilibrium~\citep{Jacobson:2011hp} are consistent with faster tidal
synchronization of the secondary, $\tau_s / \tau_p \propto q^{3/2} $
and $\tau_s / \tau_p \propto q^{5/2}$, respectively.

When the mass ratio is nearly equal, tides drive both bodies to
synchronization in nearly the same timescale~\citep{Jacobson:2011eq}.
From this configuration, where both members are tidally locked, the
BYORP effect can expand or shrink the mutual orbit to great affect.
Acting independently on each body, in addition to the YORP effect
acting on each component, BYORP can effectively transfer angular
momentum to the orbit~\citep{Taylor:2014by}.  This could lead to rapid
separation~\citep{Jacobson:2011eq}, or inward drift leading to
unstable
configurations~\citep{Bellerose:2008fo,Scheeres:2009dc,Taylor:2014by},
to gentle collisions and contact binaries
\citep{Scheeres:2007io,Jacobson:2011eq}.

Although circulating tides drive the secondary to synchronous
rotation, the secondary still has significant tidal dissipation
occurring within it due to librational tides.  The circulating theory
is inappropriate for libration since according to this theory the
tidal bulge instantaneously moves across the body.
\citet{Mignard:1979kb} developed an alternative approach that assumes
that the phase lag is proportional to the frequency of the tidal
forcing.  Here $\lambda_0$ is the characteristic spin rate at which
the body transitions from a circulation torque to the libration torque
or vice versa~\citep[where $\lambda_0$ is related to tidal lag time
  $\Delta t$ by $2 Q |\lambda_0| \Delta t = 1$;][]{Mignard:1979kb}.
In the tidal torque, $\lambda_0$ takes the place of ${|\omega-n|}$ in
the denominator Eq.~\ref{eqn:tidaltorque}.  The libration torque is
not only appropriate when the system is librating, but also when the
system is circulating slowly compared to $\lambda_0$. However, this
torque becomes inappropriate as the body begins to circulate quickly
since the tidal bulge could wrap about the body.

These two theories are actually one and the same if $\lambda=
  \omega-n$ when $\omega-n > \lambda_0$ and if $\lambda = \lambda_0$
  when $\omega-n \leq \lambda_0$, in which case $\lambda$ replaces
  $\lambda_0$ in equation 2. This approximate tidal torque can handle
  both libration and circulation for nearly circular and non-inclined
  systems~\citep{Jacobson:2011eq}.

\bigskip
\noindent
{\textbf{ 4.3 BYORP effect and Tides}}
\bigskip

The leading hypothesized formation mechanism for Group A binary
asteroids is by rotational disruption, which is observed to produce a
rapidly rotating primary and a secondary that is quickly tidally
locked---the secondary is even predicted to begin rotating more
slowly~\citep{Scheeres:2007io,Walsh:2008gk,Jacobson:2011eq}. When
considering this configuration for nearly circular orbits, circulating
tides on the primary and librational tides on the secondary contribute
to the change in eccentricity of the mutual orbit. Since the libration
of the secondary and the mutual eccentricity are
coupled~\citep{McMahon:2012uo}, the librational tides on the secondary
are often broken into direct librational and radial
components~\citep{Murray:2000th}. The sum effect of all these tides on
the mutual eccentricity is that the eccentricity is always being
damped due to the dominance of the librational tides on the secondary,
for a wide range of tidal parameters
considered~\citep{Goldreich:2009ii,Jacobson:2011hp}.

In the singly synchronous configuration---rapidly rotating primary and
tidally locked secondary, the mutual orbit of a small binary asteroid
can evolve according to both the BYORP effect and tides. While tides
in synchronous binary asteroids systems act only to expand the
semi-major axis and only to decrease eccentricity, the BYORP effect
can expand or shrink the semi-major axis depending on the shape and
orientation of the secondary. In the case of BYORP effect driven
expansion, both processes are growing the semi-major axis and both are
reducing the eccentricity. As discussed above, this process can
  lead to disruption at the Hill radius or de-synchronization of the
  secondary, which can strand the mutual orbit at a wide semi-major
  axis due. Alternatively, the BYORP effect and tides can act in
opposite directions. These effects drive the semi-major axis to an
equilibrium location:
\begin{equation}
a^* = \left( \frac{2 \pi k_{2,p} \omega_d^2 \rho R_p^2 q^{4/3} }{ B_s H_\odot Q_p} \right)^{1/7}
\end{equation}
This semi-major axis location depends directly on the tidal parameters
and the BYORP coefficient. If this location is distant, then
  secondaries could be rapidly
  lost~\citep{Cuk:2005hb,Cuk:2007gr,Goldreich:2009ii,McMahon:2010jy},
  and the binary formation rate would have to be significant to
  account for the observed $\sim15\%$
  fraction~\citep{Cuk:2007gr}. Alternatively, the proposed equilibrium
  of tides and BYORP prevents this rapid destruction of systems and no
  longer requires binary formation rates to match potentially very
  fast BYORP disruption rates~\citep{Jacobson:2014tm}.

A prediction for occupying this equilibrium is that the semi-major
axis should not be changing significantly. While measured changes in
the semi-major axis or orbital period require currently unobtainable
precision, a change in the semi-major axis does lead to a quadratic
drift in the mean anomaly~\citep{McMahon:2010jy}, which can be
measured very precisely through the timing of mutual events in
photometric light curves. A large survey has been undertaken to
examine if these drifts occur~\citep{Scheirich:2009bv}. The first
results from this survey find no drift in mean anomaly for NEA binary
(175706) 1996 FG$_3$~\citep{Scheirich:2015ez}, which may point to this
equilibrium.
\begin{figure*}
 \plotone{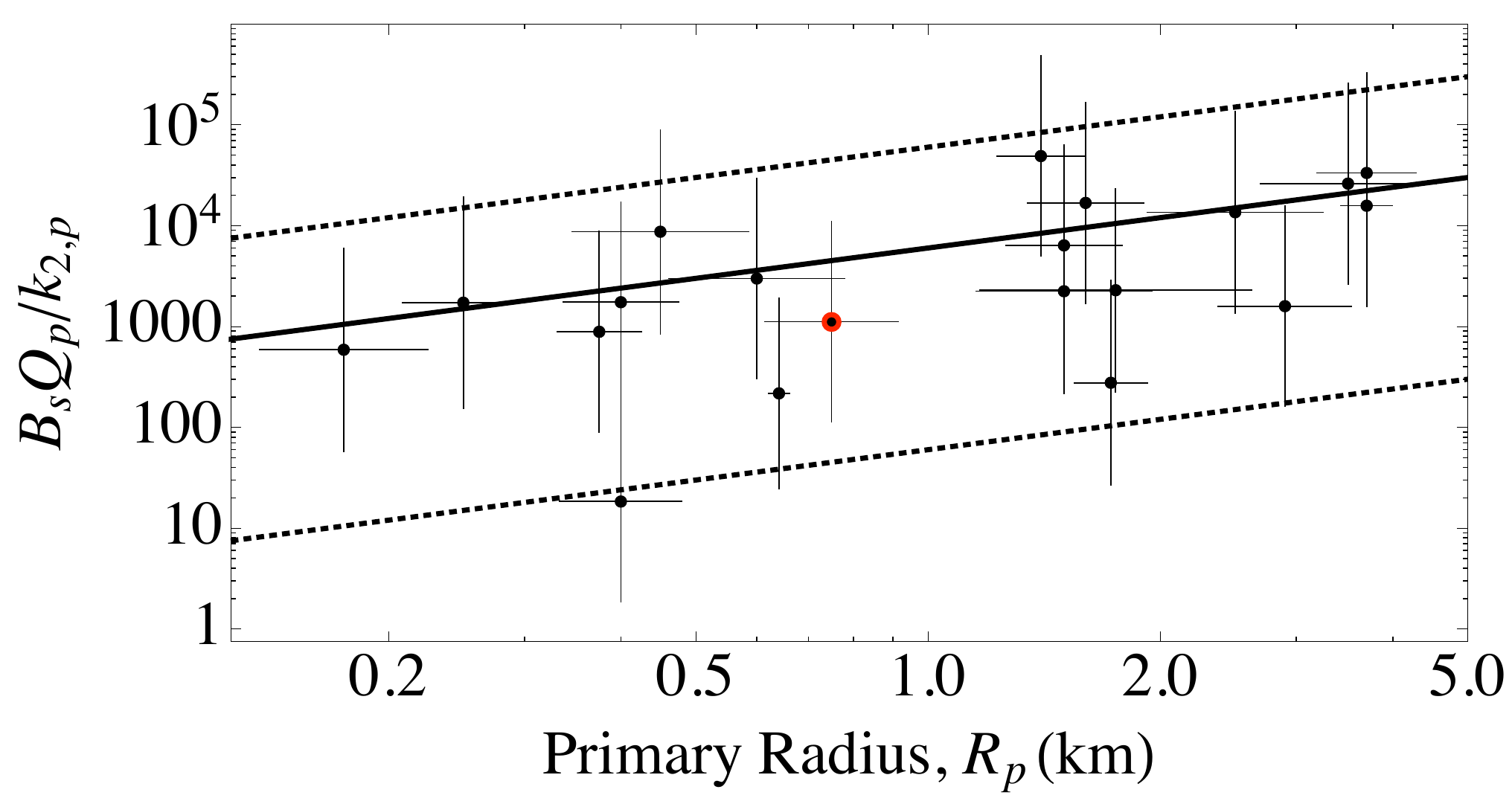}
\caption{$B Q / k_p$ were calculated directly from observed quantities
  according to equation~\ref{eqn:BQK} for each known synchronous
  binary, and plotted as a function of primary radius $R_p$ along with
  1-sigma uncertainties. A red circle highlights the binary (175706)
  1996 FG$_3$. All the data is from
  http://www.asu.cas.cz/~asteroid/binastdata.htm maintained by Petr
  Pravec according to~\citet{Pravec:2006bc}. The solid line is a
  fitted model to the data: $B_s Q_p / k_{2,p} = 6 \times 10^{3}
  R_p$. The dashed lines indicate the range of predicted scatter in
  the model due to the BYORP coefficient (possibly 10 times stronger
  or 100 times weaker). Reproduced with some updated binary parameters
  from~\citep{Jacobson:2011hp}.}
\label{fig:BQkPlot}
\end{figure*}

If the singly synchronous binary population occupies this equilibrium,
then we are able to learn about the internal properties of asteroids
from only remote sensing measurements~\citep{Jacobson:2011hp}, however
the tidal and BYORP coefficients are degenerate:
\begin{equation}
\frac{B_s Q_p}{k_{2,p}} = \frac{2  \pi \omega_d^2 \rho R_p^2 q^{4/3}}{H_\odot a^7}
\label{eqn:BQK}
\end{equation}
This parameter relationship is shown in Figure~\ref{fig:BQkPlot} along
with a fit to the data: $B_s Q_p / k_{2,p} = 6 \times 10^{3} \left(
R_p / 1\text{ km} \right)$. As discussed above, estimates for the
BYORP coefficient are around $B_s \sim
0.01$~\citep{Cuk:2005hb,Goldreich:2009ii,McMahon:2010jy,McMahon:2012ti}. From
the data, the tidal parameters then follow: $Q / k_2 = 6 \times 10^5
\left( R_p / 1\text{ km} \right)$, very different than the $Q / k_2
\gtrsim 10^7 \left( 1\text{ km} / R_p \right)$ predicted from a
modification of the continuum theory for rubble pile
asteroids~\citep{Goldreich:2009ii}.

\citet{Taylor:2011bj} predict tidal properties by assuming a tidal
evolutionary path from twice the primary radius to the current orbital
separation in under a certain timescale. For (175706) 1996 FG$_3$,
they estimate that $Q/k_2 \approx 2.7 \times 10^7$ in order to migrate
from 2 to 3.6 primary radii in 10 Myr. Using the new estimate of the
tidal parameters from~\citet{Scheirich:2015ez}, $Q/k \approx 2.4
\times 10^5$ and this same tidal migration (assuming no influence from
the BYORP effect) could take place in $5.6 \times 10^4$ years. This
much higher rate of tidal dissipation or much larger tidal Love number
can only be consistent with a tidal rheology very different than
terrestrial planets and moons. Furthermore, the equations which
convert the tidal Love number to a rigidity or elastic modulus, often
denoted $\mu$, assume a continuum model that may not apply for this
rubble pile tidal rheology.

So far, the discussed tidal theory assumes that all of the rotation
axes are aligned and that the mutual orbit is nearly circular. If this
is not the case, then the tidal bulge can have a significant effect on
the mutual orbit and rotation state of the asteroid. The differences
between the different tidal theories become more extreme, and they
differ by more than a matter of magnitudes but also of direction. This
is an ongoing area of active research with new tidal theories being
developed to eventually describe asteroid lithologies
accurately~\citep{Goldreich:2009ii,Efroimsky:2009jj}.

A separate tidal effect relies on 'tidal saltation', or the physical
lofting of material off the surface of the primary. The very rapid,
near critical, rotation of the primary permits the very small
perturbations of the secondary to loft debris off the primary's
equator~\citep{Fahnestock:2009en,Harris:2009ea}, and during flight
angular momentum is transfered from the debris to the orbit of the
secondary. This expands the orbit of the secondary at rates that could
potentially compete with tidal foces. Given the direct physical
alteration of the primary by the repeated lofting and landing of
particles on the equator, this theory provides an interesting
observational test for future observations of equatorial ridges on
NEAs.

\bigskip
\noindent
\textbf{ 4.4 Asteroid re-shaping}
\bigskip

If the asteroids were simply fluids then they would follow permissible
shape and spin configurations that have been studied by many including
Newton, Maclaurin, Jacobi, Poincar\'{e}, Roche and
Chandresekhar. Observations of asteroids clearly show that they are
not fluids and their distribution of shape and spin configurations
agree~\citep{Pravec:2002ts}. Observations also suggest that they are
not simply monolithic rocks. 

Rather, the population of small asteroids ($D<10$~km) are thought to
be primarily gravitational aggregates consisting of small bodies held
together almost strictly by their self gravity. Numerous observations
and models contribute to this line of thought, including their spin
and shape distributions but also observations of crater chains on the
Moon, the breakup of Comet Shoemaker-Levy 9, the very large observed
impact crater on the large primitive asteroid Mathilde and of course
the striking images of the small asteroid Itokawa. These arguments
were last summarized in {\it Asteroids III} by
\citet{Richardson:2002vr}, and the chapter by Scheeres et al. in this
volume reviews our general knowledge of asteroid interiors.

Efforts to understand asteroid shape and spin configurations borrow
cohesionless elastic-plastic yield criteria from soil mechanics
\citep{Holsapple:2001di,Holsapple:2004ff,Sharma:2009fo}. These
formulations calculate envelopes of allowable spin and shape
configurations as a function of the material properties---typically
relying on an angle of friction as the critical parameter.  Neither
cohesion nor tensile strength is required to explain the shapes and
spins of nearly all large ($D>200$~m) observed asteroids
\citep{Holsapple:2001di,Holsapple:2004ff}, though the spins and shapes
do not rule out any material strength either.

What about cohesion?  Multiple recalculations of allowable spin rates
as a function of cohesive forces find that even very
small amounts of cohesion can dramatically change the allowable spin
rates for a body.  Even amounts as low as 100~Pa allow for km-sized
asteroids to rotate much faster than the observed 2.3--4~hr
limit~\citep{Holsapple:2007eg,Sanchez:2014ir}. Only a single body is
observed to be larger than 200~m and rotate faster than
2.3--4~hr~\citep{Warner:2009ds}.

\citet{Rozitis:2014bx} combined measurements of Yarkovsky drift and
thermal properties to estimate the density of a km-size NEA, 1950
DA. Measurements of this asteroid's spin rate find that it is rotating
faster than what simply gravity and friction would allow, and thus it
must have non-zero cohesive strength to prevent disruption. As pointed
out by~\citet{Holsapple:2007eg} very small amounts of cohesive
strength are needed to allow bodies to rotate faster than the
classical spin limits. \citet{Hirabayashi:2014de} estimated between 40
and 210 Pa for the cohesive strength of main belt comet P/2013 R3
and~\citet{Rozitis:2014bx} estimated only $64^{+12}_{-20}$~Pa of
cohesive strength of (29075) 1950 DA. This amount of cohesion is in
line with the predictions for cohesion produced by fine grain
``bonding'' larger constituent pieces of an asteroid
\citep{Scheeres:2010ju,Sanchez:2014ir}, and would be similar to what
is found in weak lunar regolith (see chapter by Scheeres et al. in
this volume).

The exciting radar-produced shape model of 1999 KW$_{4}$ showed that
the asteroid shape held more information than could be contained in a
simple tri-axial ellipsoid model~\citep{Ostro:2006dq}. The equatorial
bulge seen in those radar shape models became ubiquitous among
primaries of other rapidly rotating asteroids (see chapter by Margot
et al.). A simple rigid ellipsoid that increases its angular momentum
will drive surface material towards its equator and this happens
before the material would simply become unbound
\citep{Guibout:2003vk}. This suggests that shape change would precede
mass-loss if there is loose material available.

Basic granular flow models can estimate what shapes the body might
actually take. As the spin rate increases, the effective slope angle
on the surface changes owing to the increased centrifugal force, and
as slopes become higher on certain regions of the surface they can
surpass critical values (angle of repose or angle of friction) and
fail. After failing, material will flow ``down'' to the potential lows
near the equator settling in at lower slopes. This model found a
surprisingly good match for the equatorial ridge shape of 1999
KW$_{4}$, using an angle of failure of 37$^\circ$
\citep{Harris:2009ea}. The failure causes very regular slopes through
mid-lattitudes on nearly-circular bodies, a trait clearly observed in
the shape of 1999 KW$_{4}$~\citep{Harris:2009ea,Sanchez:2014ir,Scheeres:2015hr}.


Discrete particle approaches to modeling rubble pile interior
structure and evolution rely on $N$-body billiard ball style granular
mechanics. Many of the first models of rubble pile dynamics, tidal
disruption and spin/shape configurations relied on hard spherical
particles that never overlap or interpenetrate
\citep{Richardson:1998bq,Richardson:2005hl}. While these
``hard-sphere'' incarnations of the models did not directly account
for friction forces,~\citet{Richardson:2005hl} showed that standard
hexagonal closest packing configurations of the spheres produce enough
shear strength so that modeled bodies can maintain spin and shape
configurations within $\sim$40$^{\circ}$ angle of friction allowable
envelopes produced by~\citet{Holsapple:2001di}. While different and
more simplistic than the ``soft-sphere'' representations used to model
rubble pile asteroids
\citep{Sanchez:2011kw,Sanchez:2012hz,Schwartz:2012jd}, the modeled
aggregates can hold shape and spin configurations similar to those
observed on actual asteroids~\citep[see][Fig.10]{Walsh:2012jt}.
  Further detail can be found in Murdoch et al., in this book.

When a rubble pile asteroid is slowly spun-up by the YORP effect it
can eventually be pushed to mass-loss
\citep{Rubincam:2000fg,Vokrouhlicky:2002cq,BottkeJr:2002tu}. If the
asteroid is made of only a very few constituent pieces than they will
reconfigure and eventually separate~\citep{Scheeres:2007kv}. What
happens to those two components is a complicated dynamical dance that
involves angular momentum transfer due to non-spherical shapes
\citep{Scheeres:2007kv,Jacobson:2011eq}.

When the asteroid is made of thousands of particles different
evolutions are found for different particle surface interactions. The
``hard-sphere'' models include dissipation of energy during
collisions, but have to rely on structural packing (crystalline) to
provide shear strength rather than surface friction
\citep{Walsh:2008gk,Walsh:2012jt}. These models found model asteroids
can maintain oblate shapes at critical rotation rates, which leads to
equatorial mass-shedding. While it was hypothesized that this could
lead to in-orbit growth of the satellite
\citep{Cuk:2007gr,Walsh:2008gk}, it is clear from the dynamics of such
close orbits~\citep{Scheeres:2009dc,Jacobson:2011eq} that to
avoid almost immediate ejection from the system that there would have
to be many particles shed at the same time in order to collide,
circularize and stabilize their orbits beyond the Roche Limit.

\citet{Sanchez:2011kw,Sanchez:2012hz} utilized ``soft-sphere''
granular models, which allow for more complex surface interactions,
including various friction forces and inter-particle cohesion. These
works explore a wider range of parameters and find a large variety of
outcomes, including ``fission'' events of bodies splitting into nearly
equal parts. There is still a strong dependence on the angle of
friction for the outcome, with some of the observed oblate shaped and
critically rotating outcomes observed.

Observed mass-loss events have been associated with YORP-induced
rotational
fission~\citep{Jewitt:2013di,Jewitt:2014fe,Sheppard:2015cw,Jewitt:2015gb}. The
rotation period of (62412) 2000 SY$_{178}$ is only 3.33
hrs~\citep{Sheppard:2015cw}. Minor components outside of the dust are
difficult to observe and the shape of the primary is impossible to
deduce. Future observations are necessary to determined whether the
dust is associated only with surface failure or satellite formation in
these cases.

\bigskip
\noindent
\textbf{ 4.5 Kozai and planetary encounters}
\bigskip

Most secondaries in NEA systems are too close to their primary to
experience excursions in eccentricity or inclination due to the Kozai
effect~\citep{Fang:2012dp}. For more distant NEA systems, such 1998
ST$_{27}$ at $a\sim$16~R$_\mathrm{pri}$, the Kozai effect could play a
role of disrupting systems or driving them to collision and possibly
making contact-binaries~\citep{Fang:2012dp}.

Binary systems in the Main Belt do not encounter planets, but those on
near-Earth orbits can have encounters with terrestrial planets close
enough to alter or disrupt their systems
\citep{Farinella:1992he,Farinella:1993tr,Walsh:2008gx,Fang:2012go}. The
timescales for encounters close enough to disrupt or disturb a system
depend on its heliocentric orbit (how frequently it approaches a
planet), and also depend strongly on the system's
properties---primarily the separation of the primary and
secondary. Disruption of a typical system with $a=4$~R$_\mathrm{pri}$,
where $a$ is semimajor axis and R$_\mathrm{pri}$ is the primary's
radius, becomes significant (50\% of encounters randomized over
phasings) at encounters of 3~R$_\mathrm{Earth}$, which occur on
average every 2~Myr for NEAs \citep{Walsh:2008gx}. Planetary flybys
may also stymie other evolutionary effects, such as BYORP, by either
torquing the secondary and breaking its synchronous rotation, or by
exciting its orbital eccentricity~\citep{Fang:2012go}. Eccentricity of
0.2 can be excited for a similar $a=4$~R$_\mathrm{pri}$ system at only
8~R$_\mathrm{Earth}$, which happen every $\sim$~1~Myr on average for
NEAs~\citep{Walsh:2008gx,Fang:2012go}.

\bigskip
\centerline{\textbf{ 5. The Future}}
\bigskip

The advances made in the last decades have been driven by the
increased database of known binary systems and the mounting evidence
and measurements of thermal effects acting widely in the Solar
System. Naturally many questions remain, and we are optimistic that
trajectory of current studies is well-aligned to answer many of the
outstanding questions. We roughly outline the expected progress,
discoveries and events that we think will be the focus of {\it
  Asteroids V} chapter on this topic in a decade.

1. More observations from a variety of sources will help to expand the
catalog of rare and outlier populations. Large-scale surveys should
provide a flood of data and continue to grow our catalog (GAIA,
LSST). While small telescopes, including significant contributions
from amateurs, have been the basis of many lightcurve discoveries of
small systems, some of the high-cadence all-sky survey telescopes may
begin to eclipse the production of the network of small telescopes.

2. The non-detection of BYORP at 1996 FG$_3$ is curious and possibly
revealing~\citep{Scheirich:2015ez}. While there is a proposed theory
to explain why and how the effect may be balanced by tides
\citep{Jacobson:2011eq}, and other non-tidal effects may be similarly
important~\citep{Fahnestock:2009en,Harris:2009ea}, a non-detection is
not a detection, and the community awaits a measurement of this
interesting thermal effect. A system with a more distant companion, or
perhaps one of the triple systems, may allow for a detection in an
environment where tides are small and BYORP is
strong~\citep{Pravec:2010tc}.  The BYORP effect may be a fundamental
and dominating mechanism that is widely shaping the observed
population of small binary asteroids---so observing it in action will
be a great step forward.

3. There are spacecraft visits planned to asteroids with
``top-shapes''.  The KW$_4$ shape, or top-shape that is becoming
ubiquitous in shape models of the primaries of binary systems was a
revealing discovery in this field. Hopefully careful mapping and
geologic studies of these systems will reveal how small asteroids take
that particular shape. In turn, knowing how the ridge formed might
help researchers answer the many remaining questions about the
formation and evolution of the satellites so often found around these
top-shaped bodies. The currently planned space missions from NASA and
JAXA, OSIRIS-REx and Hayabusa II respectively, are each currently
seeking to visit primitive NEAs, and each target appears to show some
signs of an equatorial ridge~\citep{Nolan:2013gj,Lauretta:2014hi}. It
is hoped that the mission surveys of the asteroid surface will
elucidate the re-shaping histories of these bodies by showing signs of
material flow patterns, variations in ages of different surface
features and material differences in different geologic units.

\textbf{ Acknowledgments.} KJW was partially supported by the
NASA Planetary Geology and Geophysics Program under grant NNX13AM82G. SAJ was supported by the by the European Research Council Advanced Grant ÔACCRETEÕ (contract number 290568).
\bigskip

\bibliography{biblio}
\bibliographystyle{apalike}

\end{document}